# Broadband Angular-selective Mid-infrared Photodetector

Ziwei Fan[1,2], Taeseung Hwang[3], Yixin Chen[3], Zi Jing Wong[1,2,3*]

[1] Texas A&M University, Department of Aerospace Engineering, College Station 77843, United States of America

[2] Eastern Institute of Technology, School of Electronic Science and Technology, Ningbo 315200, People's Republic of China

[3] Texas A&M University, Department of Materials Science and Engineering, College Station 77843, United States of America

*Corresponding author. E-mail: zijing@tamu.edu

## Abstract

Mid-infrared photodetectors are susceptible to background noise since every object in the surroundings emits thermal radiation from different directions. To reduce this background noise and enhance signal-to-noise ratio of mid-infrared sensing, different strategies to achieve angular-selective filtering have been proposed. However, these methods are either wavelength- and polarization-dependent, or require bulky lens or mirror systems. The former compromises the photodetector sensitivity, and the latter makes it difficult to integrate with wearable or on-chip devices. In this study, we present a novel angular-selective microstructure array that can seamlessly integrate onto a mid-infrared photodetector. Our compact device leverages the conservation of etendue to attain high angular selectivity over a broad range of mid-infrared wavelengths. Radiation from unwanted angles is substantially filtered, which leads to a markedly enhanced photodetection signal-to-noise ratio. Furthermore, the device's photoresponse is shown to be polarization- and wavelength-insensitive, avoiding signal losses associated with narrow spectral ranges or polarization dependence, and therefore circumventing degradation in photodetector sensitivity. Our broadband angular-selective mid-infrared photodetector holds great promise for wearable devices, medical diagnostics and space applications.

**Keywords**  Thermal radiation; Infrared sensing; Angular selectivity; Non-imaging optics; Broadband detector.

## Introduction

Thermal radiation conveys information about the temperature and composition of the emitter, as manifested by its intensity and spectral signatures. Therefore, the detection of thermal radiation has been widely used in remote temperature sensing (1,2), thermal imaging (3-6), infrared spectroscopy (7,8) and infrared astronomy (9-11). While mid-infrared (mid-IR) photodetectors are highly versatile in their applications, they face the challenge of constantly receiving thermal radiation from all objects with temperature above 0 K, which creates a complex thermal environment. Such unavoidable thermal radiation adds to background noise of the mid-IR photodetector and degrades the signal-to-noise ratio (SNR), a critical and fundamental criterion for photodetectors (12,13). Thermal radiation is known to be broadband and unpolarized. As a result, it is challenging to distinguish signal from background noise through wavelength or polarization selection (Fig. 1a). Nevertheless, the finite size of the object to detect leads to the occupation of a finite solid angle. This inherent characteristic inspires angular selection: by accepting only signals impinging from a specific angular range and rejecting those outside the angular range, the background noise can be significantly reduced (14) (Fig. 1b).

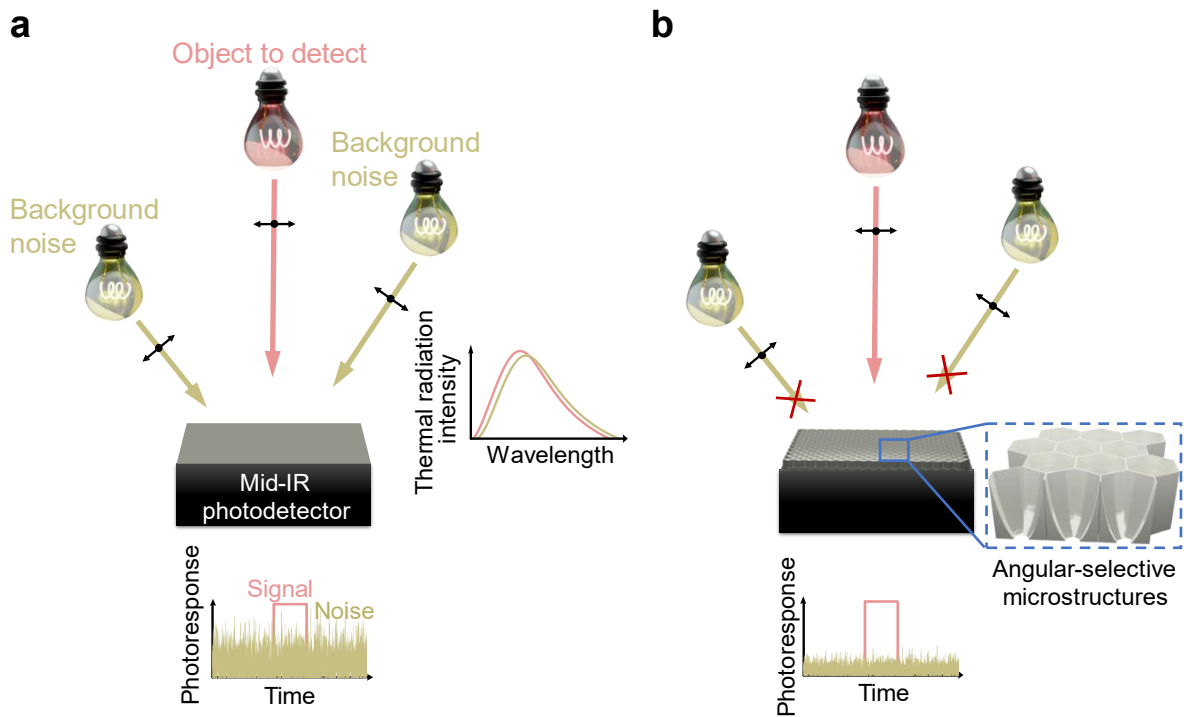

**Fig. 1** Broadband angular-selective infrared photodetector (BASIP). **a** Conventional mid-IR photodetectors inevitably receive background noise from the complex thermal environment,

significantly constraining the photodetector's signal-to-noise ratio. **b** A BASIP is realized by integrating a mid-infrared photodetector with angular-selective non-imaging microstructures, which are composed of oppositely-facing micro parabolic reflectors arranged in a hexagonal pattern. The angular-selective microstructures selectively allow thermal radiation from a specific angular range to transmit through and reach the underlying photodetector. This allows the photodetector to detect signals from the target object while blocking noise from other thermal sources in the environment. Signal-to-noise ratio (SNR) is thus enhanced. The zoomed-in structure shows the three-dimensional angular-selective microstructures designed with an acceptance angle of 15°.

While directional selectivity is often achieved with bulky lens or mirror systems (15,16), on-chip (17), wearable (18) and biomedical (19) photodetectors require compact directional-selective devices. Some progress has been made in achieving directionality with nanoscale or microscale structures. Resonant (20-22) and diffractive (23, 24) structures have been created to achieve directional selectivity. However, such strategies can only achieve directionality within a narrow angular range, significantly limiting the signal intensity of a mid-IR photodetector. Gradient epsilon-near-zero (ENZ) metamaterials were recently developed to expand the spectral range of directional selectivity to a few micrometers. Nevertheless, the gradient ENZ metamaterials still showed substantial absorption of thermal radiation at unwanted directions and weak directionality for certain in-band wavelengths (25). Another proposed approach involved plasmonic gratings that utilized directional impedance matching, but this design has not been experimentally realized due to its extreme dimensions (26-28). Alternatively, a one-dimensional photonic crystal was employed to reflect light incident from all directions other than the Brewster angle. However, the clean cutoff of excessive transmitting channels requires a large number of thin films, typically hundreds of layers, making the entire structure bulky and cost-inefficient (29,30). Furthermore, all these existing techniques exhibit polarization dependence and most of them can operate only at oblique angles. The former considerably decreases signal intensity and leads to degraded photodetector sensitivity. The latter means that the photodetector to receive thermal radiation from a wide range of directions, spanning all azimuthal angles. This creates a transmission or absorption profile in the shape of a hollow cone, which hinders the photodetector from exclusively detecting signals originating from the desired object. Hence, there is a pressing need for new techniques that enable broadband polarization-insensitive transmission solely in the normal direction with compact structures.

In this work, we create angular-selective microstructures composed of non-imaging microphotonic elements with a height of 89 μm and a period of 70 μm. These microstructures exploit the conservation of etendue and restrict the angular spread of light by expanding its areal spread. Unlike conventional bulky Winston cones used for beam condensing (31-33), our microphotonic structures are highly miniaturized and compact. Our results further demonstrate seamless integration of the angular-selective microstructures with commercial mid-IR photodetectors, which not only improves the photodetector's SNR, but also enables directional photoresponse centering at the normal direction. Moreover, we validate that the photoresponse of the angular-selective photodetector is both polarization-insensitive and broadband.

**Experimental Section**

**Simulations**

In this study, simulations were conducted using the COMSOL Multiphysics finite-element-method solver. Periodic boundary conditions were applied to the hexagonal unit cell's edges, and perfectly matched layer boundary condition were applied to the top and bottom of the simulated region. As for the parabolic reflectors, they were modeled as perfect electric conductors.

**Fabrication**

A custom setup was used to mount the commercial mid-IR photodetector (HCS, Heimann Sensor Gmbh) onto a glass substrate, ensuring its top surface remained horizontal for following steps. The details about the custom setup can be found in Supporting Section 3. The angular-selective microstructures' polymer structure was fabricated on the photodetector using three-dimensional nanolithography based on two-photon polymerization (TPP). The laser power and galvo scan speed were set to 35 mW and 10000 µm/s, respectively. Afterwards, the polymer microstructures were developed in propylene glycol monomethyl ether acetate for 20 minutes, followed by isopropanol for 5.5 minutes, and methoxynonafluorobutane for 2.5 minutes. The microstructures were then cured under ultraviolet light. Next, the mid-IR photodetector was mounted on a homemade glass sample holder for oblique-angle electron beam deposition of 400 nm thick silver layers. The deposition rate was 5 Å/s. To ensure silver coverage of the parabolic reflectors at various orientations, electron beam deposition was sequentially performed from three different azimuthal angles. The silver deposited on the bottom aperture was later removed using argon plasma etching.

Additionally, a reference photodetector was fabricated for comparison. Using TPP nanolithography, photoresist layer was patterned to match the area and shape of the angular-selective microstructures. Subsequently, silver was deposited using electron beam evaporation, after which the photoresist was carefully stripped off. This process left only the microstructure-matching area exposed to the air and the remaining region was covered by silver.

**Characterization**

Thermal sources were strategically positioned in front of the photodetector for various measurements. To measure directionality, the photodetector was positioned at a sufficient distance from a thermal source and rotated to different orientations. For the signal-to-noise ratio (SNR) characterization, three thermal sources were used, with two activated periodically. When assessing the ability of BASIP to distinguish the source of the signal, two thermal sources were used, with one activated periodically. The photoresponse of both the BASIP and the Ref-photodetector was captured using a digital multimeter (Keithley DAQ6510).

To characterize the BASIP's photoresponse versus polarization or wavelength, a polarizer or bandpass filter was positioned between the source and the photodetector. For the evaluation of the BASIP's photoresponse at different wavelengths, a specialized thermal emitter with a $CaF_2$ window was utilized. This window is highly transmissive within the 2 to 11 μm range and blocks unwanted long-wavelength thermal radiation.

## Results and Discussion

**Design of angular-selective microstructures**

We create angular-selective non-imaging microstructures composed of parabolic metallic surfaces. These angular-selective microstructures are designed to direct light from a top aperture with area $A_1$ to a much smaller bottom aperture with area $A_0$. During this process, the etendue, a quantity proportional to the product of cross-sectional area and the diverging angle of light, is conserved (34,35). As a result, the permitted angular spread at the large top aperture is restricted. Light is allowed to transmit through the angular-selective microstructures (Fig. 2a) and get absorbed by the photodetector only if its angle of incidence (AOI) falls within this restricted angular range, while other light that falls outside this angular range is reflected (Fig. 2b). The maximum angle allowed for transmission is called acceptance angle ($\theta_a$). In contrast to previous works based on resonant or propagating modes, the non-imaging microstructures emphasize the manipulation of etendue, which is not sensitive to wavelength or polarization.

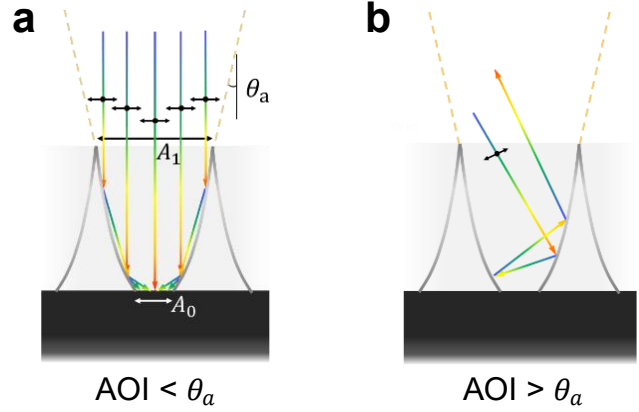

**Fig. 2** The mechanism of angular selectivity in BASIP. **a** Due to the conservation of etendue, when thermal radiation enters through the top aperture with a large area $A_1$, only light within a narrow angular range is allowed to reach the bottom aperture with a small area $A_0$. The maximum allowed angle of incidence is defined as the acceptance angle ($\theta_a$). When the angle of incidence (AOI) is smaller than the acceptance angle $\theta_a$, thermal radiation is directed to the bottom aperture. **b** When AOI > $\theta_a$, incident thermal radiation is reflected by the angular-selective microstructures and will not be detected by the photodetector.

The angular-selective microstructure unit cell (Fig. 1b) is designed to be hexagonal to enable seamless tessellation over the mid-IR photodetector surface. In this design there will be no gap between the microstructures, ensuring all incident light will be directionally controlled. Additionally, the high symmetry of the hexagonal shape facilitates polarization insensitivity even when the structure is highly miniaturized. In this study, we set the acceptance angle of angular-selective microstructures to be 15°. The influence of an angular-selective microstructure unit cell on a photodetector was examined through finite element analysis, where we modeled the photodetector as a perfect absorber. Fig. 3a-c shows the electric field distribution with light incident from different angles. At AOI of 0°, light is directed to the bottom aperture and absorbed, so strong signal intensity is predicted. When the AOI approaches 16°, close to the designed acceptance angle of 15°, the electric field is concentrated at the boundary of the bottom aperture with less light entering the photodetector. When the AOI increases to 30°, exceeding the acceptance angle, light no longer reaches the photodetector, leading to negligible photoresponse. We further present the angular-resolved photoresponse in Fig. 3d, demonstrating the evident directional selectivity induced by the microstructures. An angular range of ±15° is predicted, which is consistent with the designed acceptance angle. Additionally, Fig. 3e shows the simulated absorptance spectra of the microstructure-integrated photodetector at different AOIs, which are nearly flat, indicating uniform directional control for different wavelengths. A detailed discussion about the spectrum profile at 0° and cut-off wavelength of the BASIP is provided in Supporting Section 1. In addition to its strong directionality, the angular-selective microstructure's flexibility makes it stand out. Its angular range can be easily adjusted through structural design and angular-selective microstructures with different acceptance angles are demonstrated in Fig. S1. Also, angular-selective microstructures for detecting signals incident from off-normal directions are demonstrated in Supporting Section 2.

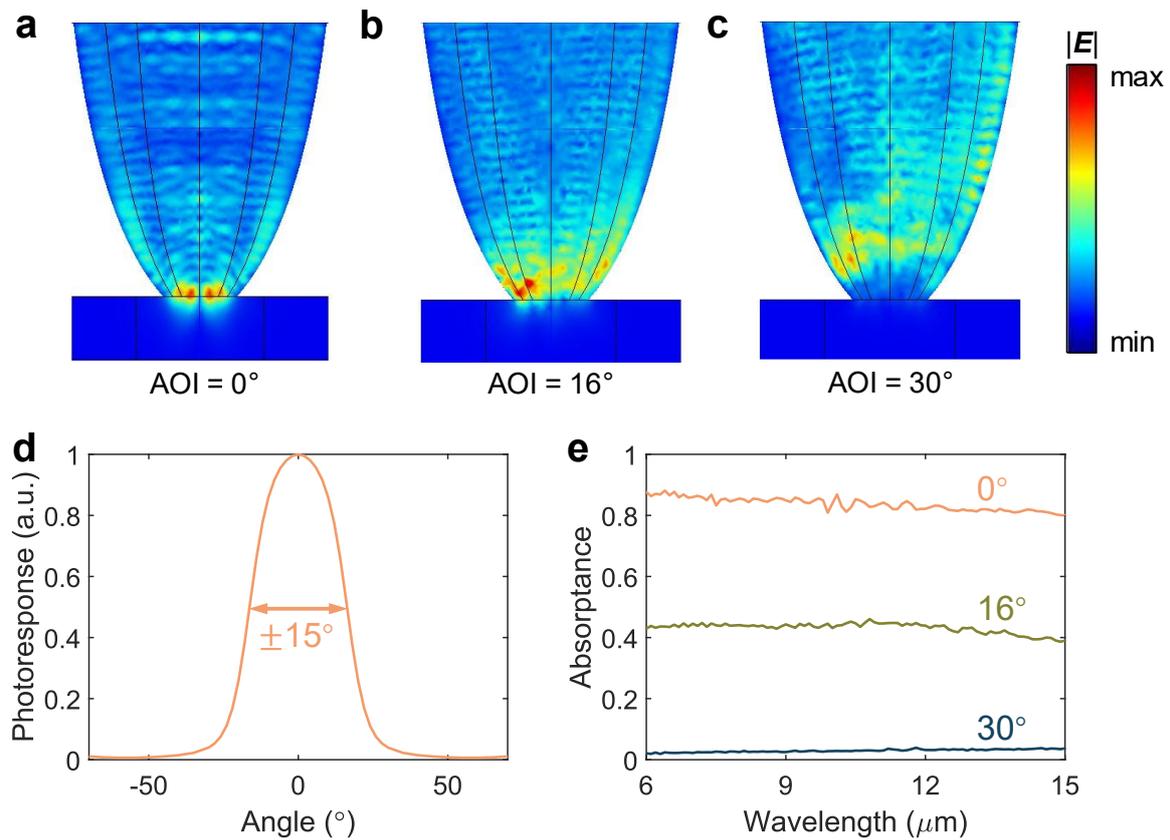

**Fig. 3** Design and simulation of angular photoresponse of the BASIP. **a-c** Electric field distribution in an angular-selective microstructure unit cell at different angles of incidence (AOIs). **a** When AOI equals 0°, incident thermal radiation is guided to the center of the bottom aperture and absorbed by the underlying absorber. **b** When AOI increases to 16°, the light field is concentrated at the edge of the bottom aperture, resulting in less light reaching the absorber.

**c** When AOI is 30°, which exceeds the designed acceptance angle, light no longer reaches the bottom aperture. **d** Simulated angular-resolved photoresponse at a wavelength of 10 μm, showing an angular width of ±15°, with photoresponse diminishing rapidly outside this angular range. **e** Simulated absorptance spectra of the angular-selective microstructures and underlying photodetector at different AOIs. The absorptance remains nearly constant across wavelengths for a given AOI. However, when the AOI exceeds the designed 15° acceptance angle, the absorptance drops dramatically.

**Realization of angular-selective microstructures**

The angular-selective microstructures were directly fabricated onto a commercial mid-IR photodetector using two-photon polymerization (TPP) nanolithography. Subsequently, a conformal coating of silver (Ag) was applied to the parabolic surfaces via oblique-angle deposition. During this process, the top part of angular-selective microstructures acted as a mask, preventing Ag from reaching the underlying photodetector. To ensure that the photodetector remains free of silver coverage, Ar plasma etching was carried out. The scanning electron microscope (SEM) image in Fig. 4a shows well-defined and uniformly coated structures. The transparency of the bottom aperture was verified through the apparent absence of Ag at that position. As shown in Fig. 4b, the thickness of the angular-selective microstructures after lithography was negligible compared with the thickness of the photodetector, and the microstructures were significantly smaller than conventional mirror or lens systems.

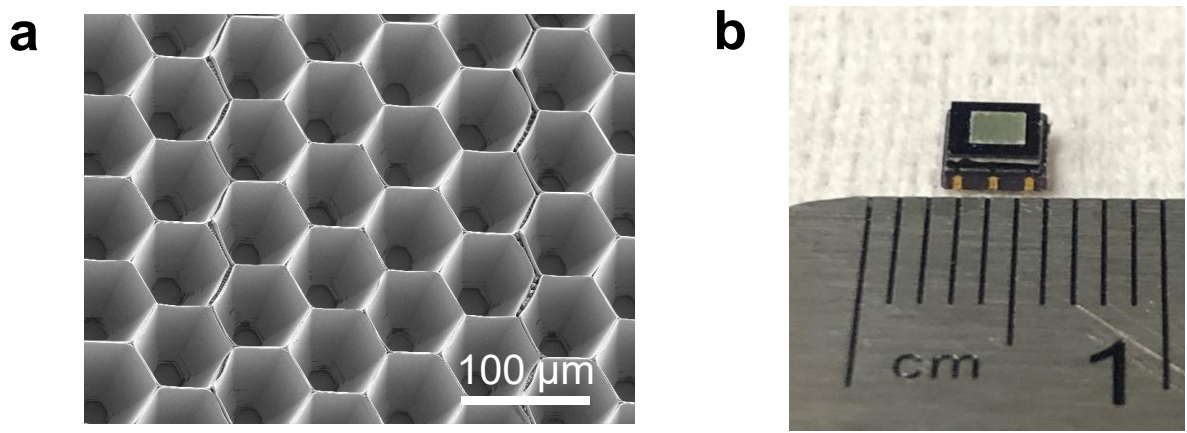

**Fig. 4** Microstructure fabrication and integration. **a** SEM image of the angular-selective microstructures, showing high structural quality. The bottom aperture is clearly free of Ag coverage. **b** Photograph of the BASIP where angular-selective microstructures were directly

fabricated on a mid-infrared photodetector through two-photon polymerization lithography. The ruler's unit is centimeters. This photo was taken before the deposition of silver to clearly demonstrate the position and miniaturized dimensions of the microstructures.

**Strong directionality of the BASIP**

To experimentally prove the directional selectivity of the BASIP, a thermal source with diameter $d$ of 3 mm was positioned in front of the BASIP (Fig. 5a). The temperature of the thermal source was set at 450 °C and its distance $D$ from the BASIP was significantly larger than the diameter of the source ($D = 40$mm $\gg d$). We rotated the BASIP to different directions and its angular-resolved photoresponse was recorded in Fig. 5b. The BASIP demonstrates evident directional selectivity: a full width half maximum (FWHM) of $\pm 15°$ was achieved, which agrees well with the simulation. On contrary, the reference photodetector (Ref-photodetector) with an identical active area as the BASIP, showed a wide FWHM of $\pm 55°$, nearly 3 times broader than that of the BASIP.

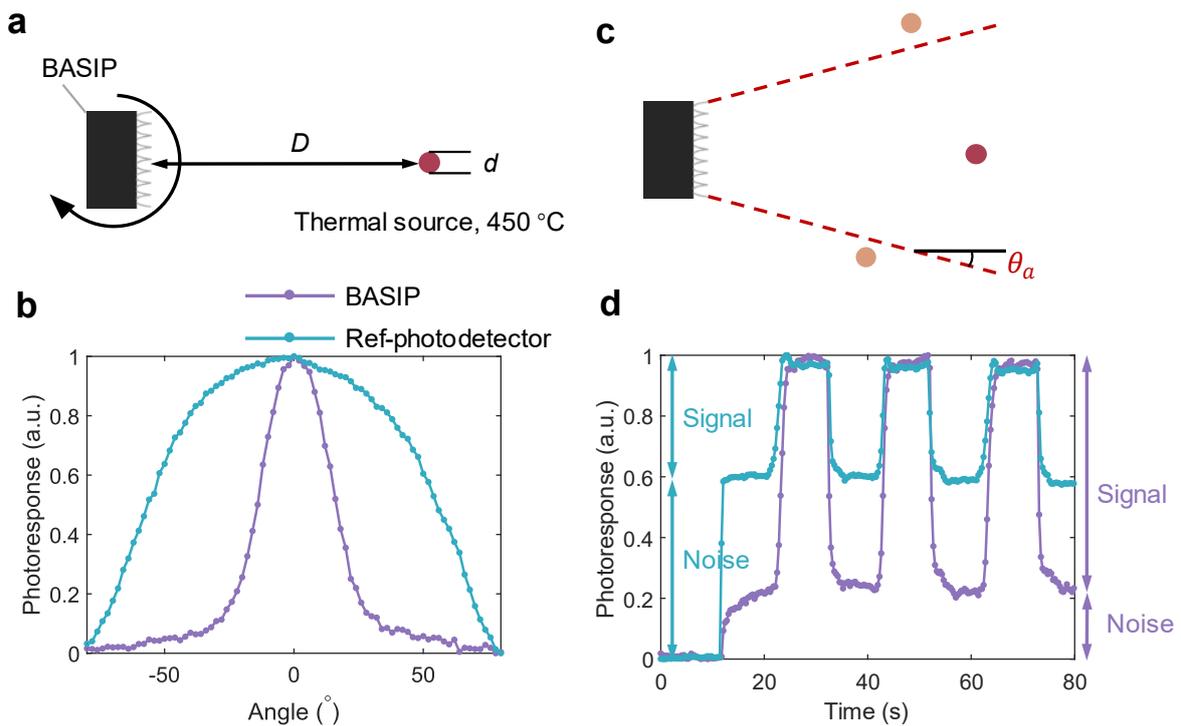

**Fig. 5** Characterization of directional selectivity and signal-to-noise ratio (SNR). **a** To characterize the directional selectivity, a thermal source with diameter $d$ was positioned at a distance $D$ from the photodetector ($D \gg d$). The photodetector was rotated to different orientations. **b** Experimentally measured angular-resolved photoresponse of the BASIP and reference photodetector (Ref-photodetector) without microstructures. The BASIP exhibits a much narrower angular range ($\pm15°$) compared to the Ref-photodetector ($\pm55°$), verifying the strong directional selectivity of the BASIP. **c** Illustration of the experimental setup for SNR measurement. Three thermal sources were used: one (red) positioned within the BASIP's angular range, while the other two (orange) laid outside this angular range. **d** Initially, all sources were off. At the 10$^{th}$ second, the orange sources were turned on, introducing background noise. After the 20$^{th}$ second, the red source was activated periodically, generating the signals. In the Ref-photodetector, the noise was much stronger than the signal, resulting in an SNR of 0.65. In contrast, the BASIP significantly suppressed the noise, yielding an enhanced SNR of 3.92.

**Enhanced SNR of the BASIP**

More experiments were conducted to compare the SNR of the BASIP and the Ref-photodetector. Three thermal sources were placed in front of the photodetector, with one within the acceptance angle (the red source in Fig. 5c) and two outside the acceptance angle of the BASIP (the orange sources). Initially, all the thermal sources were turned off, and the orange ones were turned on at the 10$^{th}$ second. Subsequently, the red source was activated and deactivated periodically. The photoresponse variation with time was recorded and depicted in Fig. 5d. The Ref-photodetector showed noise stronger than signal because it received IR photons emitted from all three sources. In contrast, the BASIP exhibits a significantly stronger signal compared to noise, as it effectively blocked the background noise, making the signal more distinct. The SNR was calculated using the formula SNR = $PR_{signal}/PR_{noise}$, where $PR_{signal}$ and $PR_{noise}$ represents photoresponse generated by signal and background noise thermal radiation, respectively. The SNR of the BASIP is calculated to be 3.92, showing a significant improvement over the Ref-photodetector's SNR of 0.65.

To further exploit the directionality of the BASIP, we demonstrate its ability to distinguish the origin of signals (Fig. S2). Two sources numbered 1 and 2 were positioned in front of a photodetector. Source 1, generating true signals, fell within the accepted angular range of the BASIP, while source 2 which generated interrupting signals lied out of the accepted angular range (Fig. S2a). Each time one of the two sources was randomly chosen, activated for 10 s and then deactivated. This process was repeated for 5 times, generating 5 signals. Due to its directionality, the BASIP received strong signals from source 1 but weak signals from source 2. Therefore, from the photoresponse of the BASIP, it was easily recognized that the second and fifth peaks were triggered by source 1 (Fig. S2b) and others were triggered by source 2. In contrast, the Ref-photodetector's photoresponse for all signals were almost identical (Fig. S2c) and indistinguishable.

**Wavelength and polarization insensitivity of the BASIP**

To investigate the polarization and wavelength insensitivity of the angular-selective microstructures, we conducted additional characterizations. A polarizer was placed between the source and the BASIP. The polarization direction and the rotation angle of the BASIP were varied. Fig. 6a shows the photoresponse at different polarizations and AOIs. Notably, the photoresponse remained almost unchanged when the polarization direction varied, confirming the polarization-insensitive of the BASIP. In addition, the photoresponse's rapid decrease with angle confirmed the BASIP's pronounced directional selectivity. To demonstrate the wavelength insensitivity, we utilized a mid-IR bandpass filter and a thermal emitter equipped with a $CaF_2$ window, ensuring that the thermal source emitted only within a narrow spectral range. This thermal emitter was heated to a high temperature of 630 °C to enhance the emitting power and compensate the reduction in photoresponse caused by the bandpass filters. By varying the central wavelength of the bandpass filters, the photoresponse at different wavelengths was obtained. To account for the influence of blackbody radiation dispersion on the photoresponse, we compared the photoresponse ratios (PRRs) between the BASIP and the Ref-photodetector at various wavelengths. For clarity, we normalized the PRRs, as shown in Fig. 6b. The results indicate consistent values of 0.94, 0.95, and 1 for wavelengths of 7, 8, and 9 μm, respectively. These findings confirm that the BASIP exhibits wavelength insensitivity. A discussion about the slight increase of normalized PRR at 9 μm is provided in Supporting Section 4.

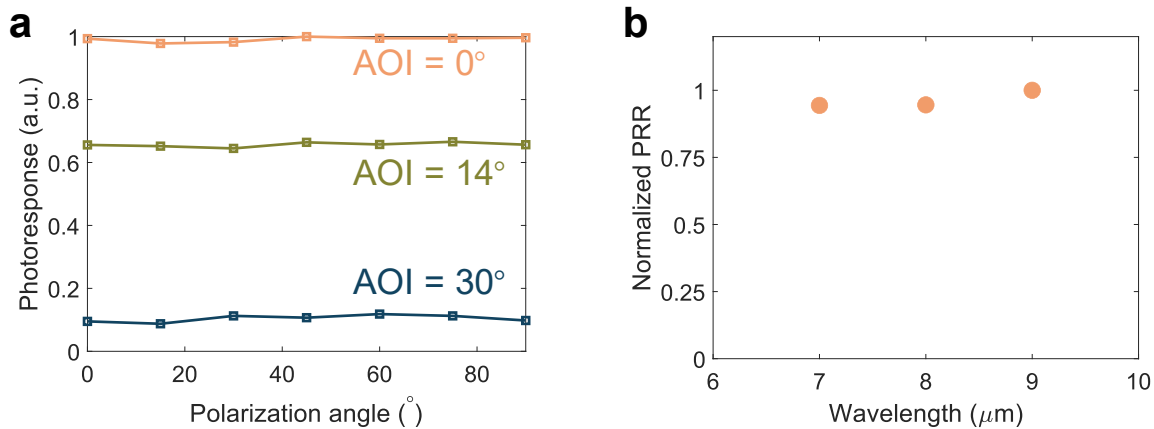

**Fig. 6** Polarization and wavelength insensitivity. **a** Photoresponse of the BASIP at various polarization directions and angles of incidence (AOIs). At a specific AOI, the photoresponse remains consistent across different polarization angles, confirming the polarization insensitivity of the BASIP. **b** Normalized photoresponse ratios (PRRs) between the BASIP and Ref-photodetector at different wavelengths, measured using bandpass filters. PRRs at wavelengths of 7, 8 and 9 μm were measured to be 0.94, 0.95 and 1, respectively, verifying the wavelength insensitivity of the BASIP.

## Conclusions

In conclusion, we developed a compact broadband angular-selective infrared photodetector (BASIP) by seamlessly integrating angular-selective non-imaging microstructures with a commercial mid-IR photodetector. The BASIP exhibited a narrow accepted angular range of ±15° centering at the normal direction, significantly smaller than the Ref-photodetector's angular range of ±55°, proving the strong directionality of the BASIP. Background noise from unwanted directions was therefore effectively suppressed, leading to a substantially improved signal-to-noise ratio. Additionally, the BASIP demonstrates polarization and wavelength insensitivity, ensuring minimal detection sensitivity degradation. Its compact design and superior performance render the BASIP highly suitable for applications in wearable devices, medical diagnostics, robotics, and space-based sensing.

**Funding**
This work was supported by the President's Excellence Fund (X-Grant) and Ningbo Yongjiang Talent Program 2022A-241-G.

**Notes**
  The authors declare no interest conflict. They have no known competing financial interests or personal relationships that could have appeared to influence the work reported in this paper.

**Supporting Information**
The following supporting information is available free of charge via the Internet at http://pubs.acs.org:
The Supporting Information includes simulated photoresponses of different angular-selective microstructures (Fig. S1), and a method to distinguish real signals from interrupting signals (Fig. S2). It also provides details on the cut-off wavelength of the angular-selective microstructures (Section 1), designs of angular-selective microstructures for detection of off-normal signals (Section 2), the detector leveling procedure before fabrication (Section 3), and analysis of the slight PRR variation at 9 μm (Section 4). (PDF)

## References

(1) Lin, W.-L., et al., Apply Iot Technology to Practice a Pandemic Prevention Body Temperature Measurement System: A Case Study of Response Measures for Covid-19. *International Journal of Distributed Sensor Networks* **2021,** *17* (5), 15501477211018126.
(2) Costanzo, S.; Flores, A., A Non-Contact Integrated Body-Ambient Temperature Sensors Platform to Contrast Covid-19. *Electronics* **2020,** *9* (10), 1658.
(3) Hu, R., et al., Thermal Camouflaging Metamaterials. *Materials Today* **2021,** *45*, 120-141.
(4) Li, M., et al., Manipulating Metals for Adaptive Thermal Camouflage. *Science Advances* **2020,** *6* (22), eaba3494.
(5) Ring, E. F. J.; Ammer, K., Infrared Thermal Imaging in Medicine. *Physiological Measurement* **2012,** *33* (3), R33.
(6) So, S.; Lee, D.; Badloe, T.; Rho, J., Inverse Design of Ultra-Narrowband Selective Thermal Emitters Designed by Artificial Neural Networks. *Optical Materials Express* **2021,** *11* (7), 1863-1873.
(7) Dorling, K. M.; Baker, M. J., Rapid Ftir Chemical Imaging: Highlighting Fpa Detectors. *Trends in Biotechnology* **2013,** *31* (8), 437-438.
(8) Hebert, G. N.; Odom, M. A.; Bowman, S. C.; Strauss, S. H., Attenuated Total Reflectance Ftir Detection and Quantification of Low Concentrations of Aqueous Polyatomic Anions. *Analytical Chemistry* **2004,** *76* (3), 781-787.
(9) Gáspár, A., et al., Spatially Resolved Imaging of the Inner Fomalhaut Disk Using Jwst/Miri. *Nature Astronomy* **2023**.
(10) Roberts-Borsani, G., et al., The Nature of an Ultra-Faint Galaxy in the Cosmic Dark Ages Seen with Jwst. *Nature* **2023**.
(11) Tabone, B., et al., A Rich Hydrocarbon Chemistry and High C to O Ratio in the Inner Disk around a Very Low-Mass Star. *Nature Astronomy* **2023**.
(12) Li, S., et al., Ultrasensitive, Superhigh Signal-to-Noise Ratio, Self-Powered Solar-Blind Photodetector Based on N-Ga2o3/P-Cuscn Core–Shell Microwire Heterojunction. *ACS Applied Materials & Interfaces* **2019,** *11* (38), 35105-35114.
(13) Shin, W., et al., Improved Signal-to-Noise-Ratio of Fet-Type Gas Sensors Using Body Bias Control and Embedded Micro-Heater. *Sensors and Actuators B: Chemical* **2021,** *329*, 129166.
(14) Shen, Y., et al., Broadband Angular Selectivity of Light at the Nanoscale: Progress, Applications, and Outlook. *Applied Physics Reviews* **2016,** *3* (1), 011103.
(15) Wilson, R. N., *Reflecting Telescope Optics Ii: Manufacture, Testing, Alignment, Modern Techniques*. Springer Science & Business Media: 2013.
(16) Ananthakrishnan, S., The Giant Meterwave Radio Telescope. *Journal of Astrophysics and Astronomy* **1995,** *16* (Sup), 427-435.
(17) Assefa, S.; Xia, F.; Vlasov, Y. A., Reinventing Germanium Avalanche Photodetector for Nanophotonic on-Chip Optical Interconnects. *Nature* **2010,** *464* (7285), 80-84.
(18) Cai, S., et al., Materials and Designs for Wearable Photodetectors. *Advanced Materials* **2019,** *31* (18), 1808138.
(19) Chow, P. C. Y.; Someya, T., Organic Photodetectors for Next-Generation Wearable

Electronics. *Advanced Materials* **2020,** *32* (15), 1902045.
(20) Jang, J., et al., Planar Optical Cavities Hybridized with Low-Dimensional Light-Emitting Materials. *Advanced Materials* **2023,** *35* (4), 2203889.
(21) Zhang, X., et al., Controlling Thermal Emission by Parity-Symmetric Fano Resonance of Optical Absorbers in Metasurfaces. *ACS Photonics* **2019,** *6* (11), 2671-2676.
(22) Costantini, D., et al., Plasmonic Metasurface for Directional and Frequency-Selective Thermal Emission. *Physical Review Applied* **2015,** *4* (1), 014023.
(23) Greffet, J.-J., et al., Coherent Emission of Light by Thermal Sources. *Nature* **2002,** *416* (6876), 61-64.
(24) Janonis, V., et al., Investigation of N-Type Gallium Nitride Grating for Applications in Coherent Thermal Sources. *Applied Physics Letters* **2020,** *116* (11).
(25) Xu, J.; Mandal, J.; Raman, A. P., Broadband Directional Control of Thermal Emission. *Science* **2021,** *372* (6540), 393-397.
(26) Argyropoulos, C., et al., Broadband Absorbers and Selective Emitters Based on Plasmonic Brewster Metasurfaces. *Physical Review B* **2013,** *87* (20), 205112.
(27) Alù, A.; D'Aguanno, G.; Mattiucci, N.; Bloemer, M. J., Plasmonic Brewster Angle: Broadband Extraordinary Transmission through Optical Gratings. *Physical Review Letters* **2011,** *106* (12), 123902.
(28) Le, K. Q., et al., Broadband Brewster Transmission through 2d Metallic Gratings. *Journal of Applied Physics* **2012,** *112* (9).
(29) Shen, Y., et al., Optical Broadband Angular Selectivity. *Science* **2014,** *343* (6178), 1499-1501.
(30) Qu, Y.; Pan, M.; Qiu, M., Directional and Spectral Control of Thermal Emission and Its Application in Radiative Cooling and Infrared Light Sources. *Physical Review Applied* **2020,** *13* (6), 064052.
(31) Ng, D. K. T., et al., Considerations for an 8-Inch Wafer-Level Cmos Compatible Aln Pyroelectric 5–14 Mm Wavelength Ir Detector Towards Miniature Integrated Photonics Gas Sensors. *Journal of Microelectromechanical Systems* **2020,** *29* (5), 1199-1207.
(32) Ng, D. K. T., et al., Miniaturized Co2 Gas Sensor Using 20% Scaln-Based Pyroelectric Detector. *ACS Sensors* **2022,** *7* (8), 2345-2357.
(33) Baulsir, C. F.; Simler, R. J., Design and Evaluation of Ir Sensors for Pharmaceutical Testing. *Advanced Drug Delivery Reviews* **1996,** *21* (3), 191-203.
(34) Winston, R.; Jiang, L.; Ricketts, M., Nonimaging Optics: A Tutorial. *Advances in Optics and Photonics* **2018,** *10* (2), 484-511.
(35) Fan, Z., et al., Directional Thermal Emission and Display Using Pixelated Non-Imaging Micro-Optics. *Nature Communications* **2024,** *15* (1), 4544.

# For Table of Contents Use Only

## Broadband Angular-selective Mid-infrared Photodetector

Ziwei Fan, Taeseung Hwang, Yixin Chen, Zi Jing Wong[*]

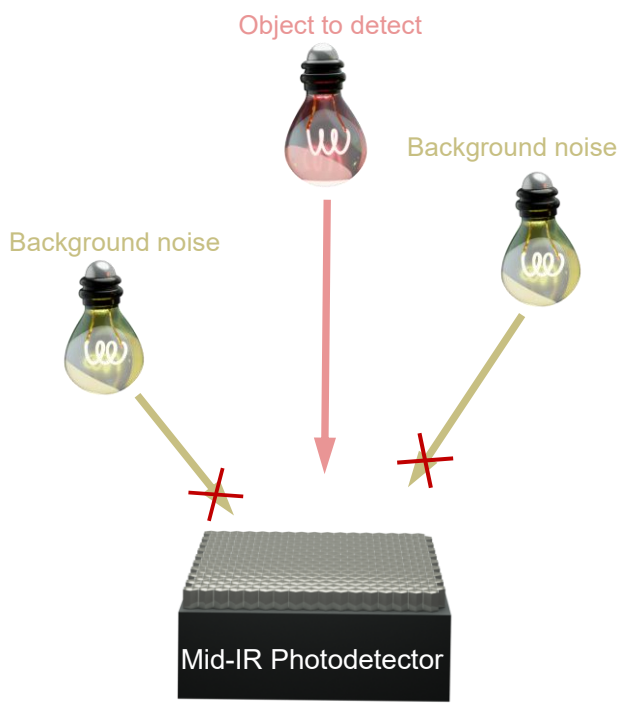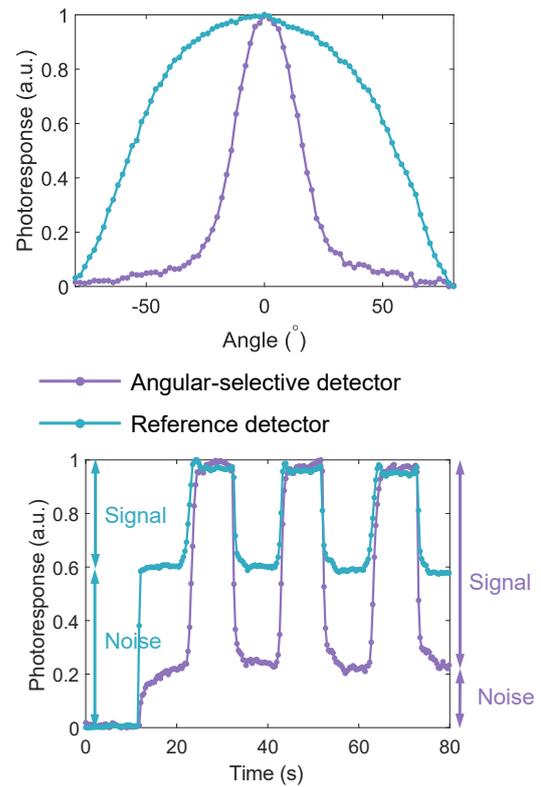

A compact mid-infrared photodetector integrated with angular-selective microstructures is created to block background thermal noise emitted from unwanted directions while detecting an object within its detectable angular range. This directional filtering significantly enhances the signal-to-noise ratio of the detector.